\documentclass[aps,pra,twocolumn,superscriptaddress]{revtex4}
\usepackage{graphicx}

\begin{document}

%\preprint{}

\title{Transverse laser modes in Bose-Einstein condensates}


\author{Graeme Whyte}
\affiliation{Department of Physics and Astronomy, University of Glasgow, Glasgow G12~8QQ, United~Kingdom}

\author{Patrik \"Ohberg}
\affiliation{Department of Physics, University of Strathclyde, Glasgow G4~0NG, United~Kingdom}

\author{Johannes Courtial}
\affiliation{Department of Physics and Astronomy, University of Glasgow, Glasgow G12~8QQ, United~Kingdom}
\email[]{j.courtial@physics.gla.ac.uk}
%\thanks{}


\date{\today}

\begin{abstract}
We examine the Bose-Einstein-condensate (BEC) equivalent of transverse aspects of laser resonators.  We model numerically repeated focussing of a 2-dimensional BEC, which could be achieved in practice by a series of far off-resonant light pulses.  We show for a range of non-linear coefficients that such a series of light pulses traps the BEC.  We also model a combination of repeated focussing and loss, which could be achieved with a series of light pulses, some far off-resonant, some resonant. In analogy to light repeatedly being focussed and locally absorbed by passing through a laser resonator, a small proportion (in our model up to 10\%) of the BEC is shaped into Hermite-Gaussian-like modes; the remainder of the BEC is lost.  This happens irrespective of the presence of a harmonic trap.  We show that astigmatic focussing of the resulting Hermite-Gaussian-like modes, which can be achieved by a pair of off-resonant light pulses, results in a specific number of vortices.
\end{abstract}

%\pacs{}
%\keywords{}

\maketitle

\section{Introduction}

Bose-Einstein condensates (BECs) are the matter-wave equivalent of lasers \cite{Rolston-Phillips-2002}:  all the atoms in a BEC are in the same quantum state, just like all the photons in a laser beam.  While in the creation of BECs this property is usually ensured by the fundamental properties of very cold bosons, in a laser it is due to the longitudinal properties of the laser resonator \cite{Siegman-1986}.  Just like the number of photons can be coherently amplified in a laser, the number of atoms can be coherently amplified in a BEC~\cite{Inouye-et-al-1999,Kozuma-et-al-1999,Chikkatur-et-al-2002}.

\begin{figure}
\includegraphics{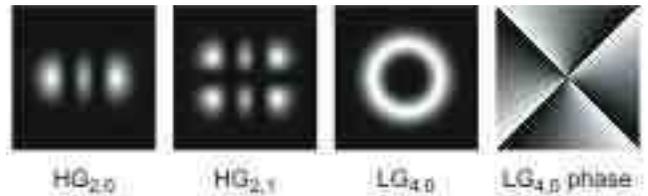}
\caption{Hermite-Gaussian (HG) and Laguerre-Gaussian (LG) modes.  Shown here are the intensity cross-sections of two HG modes, $\rm{HG}_{2,0}$ and $\rm{HG}_{2,1}$, and the intensity and phase (grayscale representation) cross-sections of one LG mode,  $\rm{LG}_{4,0}$, which possesses a central charge-4 vortex.}
\label{laser-modes-figure}
\end{figure}

Laser resonators also have an effect on the transverse structure of laser light.  The origin of this transverse structure has a huge literature devoted to it (for an overview see, for example, references \cite{Siegman-1986-laser-modes,Siegman-2000}), and a good understanding and control has proved interesting (see, for example, the discovery of fractal modes \cite{Karman-et-al-1999}) as well as useful.  For example, the standard families of laser modes, Hermite-Gaussian (HG) and Laguerre-Gaussian (LG) modes (figure \ref{laser-modes-figure}), have elegant mathematical properties (for example, they are structurally stable, i.e.\ they do not change shape on propagation, and form complete orthogonal sets of modes), and LG modes, because of their central vortex of arbitrary topological charge, are one of the most popular choices for optical experiments with optical vortices.  Most importantly, however, certain 'stable' resonators keep the light in the proximity of the optical axis -- transversally, they trap the light.

We examine here -- in the limit of the Gross-Pitaevskii equation -- a process in 2-dimensional BECs that is analogous to the formation of transverse laser modes:  periodic focussing and spatial filtering.  We believe this could be achieved experimentally with a periodic series of light pulses.  We examine this scenario for BECs in a quadratic trap and untrapped BECs.  We demonstrate the emergence of HG-like modes, and convert these into LG-like modes using the equivalent of the Gouy phase shift.  Our main concern, however, is the question if a series of light pulses could act as a novel type of trap for BECs.

\section{Eigenmode selection in laser resonators}

\begin{figure}
\includegraphics{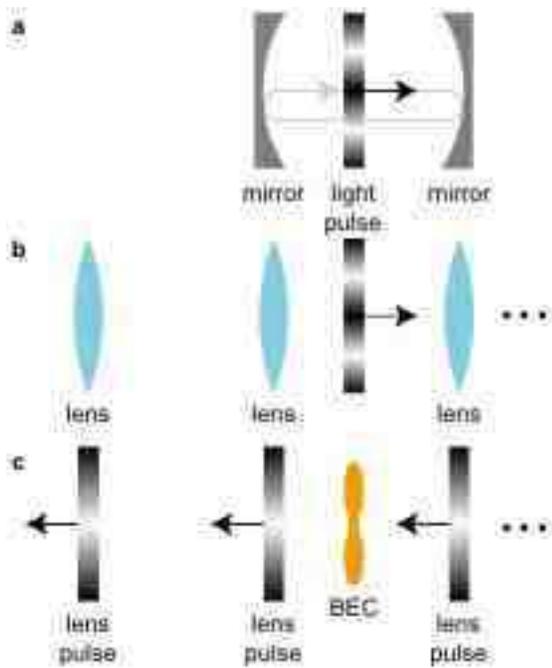}
\caption{Laser resonator~(\textbf{a}) and its unfolded lens-series equivalent~(\textbf{b}).  A light beam passing through the lens-series equivalent is periodically being focussed, just like light in the resonator is being focussed every time it is reflected off one of the resonator's mirrors.  A series of 'lens pulses' -- short, off-resonant light pulses with a specific intensity profile -- can repeatedly focus a BEC in a similar way~(\textbf{c}).  Note the role reversal:  in \textbf{a} and \textbf{b}, a light pulse is being focussed (by mirrors and lenses, respectively);  in \textbf{c} light pulses focus the BEC.}
\label{resonator-figure}
\end{figure}

Figure \ref{resonator-figure}\textbf{a} sketches one round trip of a light pulse (or alternatively a 'slice' of a cw light beam) through a standard laser resonator.  Most importantly, the light pulse is reflected back by the two mirrors facing each other, so that after one round trip, which involves reflecting off each mirror once, the light pulse is back in the same plane where it started.  One round trip is much more complicated than just some changes in propagation direction:  as the mirrors are curved, each mirror reflection also focusses the light.  In addition, localised loss occurs in one or several planes, for example through apertures or specks of dust on mirrors.  Some light beams -- the resonator's eigenmodes -- are unchanged (apart from a uniform change in intensity or phase) after one round trip through a resonator.  In the case of geometrically stable resonators, which we consider here, the eigenmodes can be divided into families (like the Hermite-Gaussian (HG) modes and the Laguerre-Gaussian (LG) modes), whereby each family forms a complete orthogonal set, i.e.\ every light beam can, for example, be described as a superposition of HG modes.  As already mentioned, the HG and LG modes can take on very recognisable shapes and have interesting features like high-charge vortices (see figure~\ref{laser-modes-figure}).

Experimentally, a pure resonator eigenmode is created in a laser by making that eigenmode the lowest-loss eigenmode, i.e.\ by ensuring the round-trip loss of all other eigenmodes is higher than that of the desired eigenmode.  This can be done, for example, by inserting absorbing cross-wires into the resonator at positions where the desired eigenmode is darker than the competing eigenmodes.  For example, an absorbing wire in the central horizontal dark line of the $\rm{HG}_{2,1}$ mode shown in figure \ref{laser-modes-figure} would not significantly increase that eigenmode's round-trip loss, but it would increase that of the $\rm{HG}_{2,0}$ mode.  The fraction of the power in the undesired eigenmodes then falls of exponentially with round trip number, after a few round trips leaving only the pure lowest-loss eigenmode.

\section{Creation of Hermite-Gaussian-like modes in BECs}

As far as transverse eigenmode selection is concerned, a resonator can be seen as a series of lenses separated by the length of the resonator (figure~\ref{resonator-figure}\textbf{b}).  If the focal lengths of the lenses match those of the resonator's mirrors, and if the important localised loss is the same, then an infinitely long lens-series equivalent of a resonator has the same eigenmodes \cite{Siegman-1986}.  Such an infinitely long lens-series equivalent of a resonator represents infinitely many round trips.  On the other hand, eigenmode selection mostly happens during the first few round trips through the resonator, which can be represented by a few lenses.  We examine here the BEC analog of such a series of lenses.

The BEC analog of an optical resonator's lens-series equivalent can consist of repetitions of lens pulses (focal time $f$) and absorption events (figure \ref{resonator-figure}\textbf{c}; absorption events are not shown).    The delay between two repetitions, $L$, corresponds to the length of the resonator.  We refer to such a configuration as a \textit{light-pulse resonator}.  We restrict ourselves to light-pulse resonators in which the absorption events always follows immediately after the lens pulse.

\begin{figure}
\includegraphics{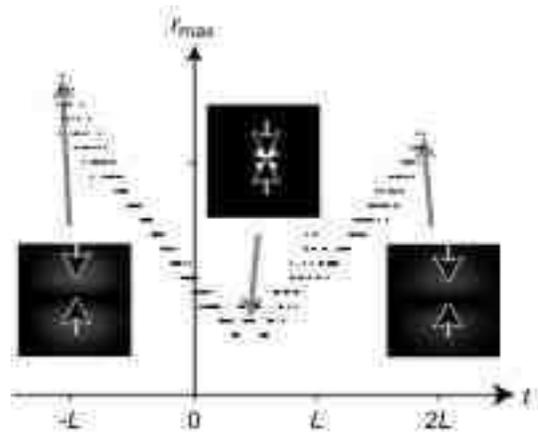}
\caption{Free time evolution of an eigenmode, calculated for $g = 0$.  If -- after the eigenmode has formed -- the series of light pulses is switched off at $t = 0$ (the last few pulses arrive at $t = -2L, -L, 0$), the BEC approximately retains its shape but changes its size.  The graph shows the distance from the centre of the maximum of the density, which indicates the relative size; the points do not lie on a smooth curve due to numerical errors.  In analogy with optical higher-order Hermite-Gaussian modes, the BEC passes through a 'focus', where it is smallest.  On either side of the focus, its size changes approximately symmetrically.  The data points for $t < 0$ were calculated using 'backward propagation' in time.  The graph is calculated for $L = 0.2$ (in adimensional time units); the vertical range then represents 3 adimensional length units.  To avoid numerical artefacts in obtaining this graph, the BEC was simulated over an area larger than the standard $16 \times 16$ non-dimensional units.}
\label{free-evolution-figure}
\end{figure}

All the pictures of BECs in light-pulse resonators shown in this paper (with the exception of two frames in figure \ref{free-evolution-figure}) are snapshots taken half-time between two lens pulses, i.e.\ a time $L/2$ after the previous lens pulse.  We refer to the sequence of events acting on the BEC between $L/2$ after the previous lens pulse and $L/2$ after the next lens pulse -- time evolution over a time $L/2$, interaction with a lens pulse, absorption event, and another time evolution through $L/2$ -- as one period of the resonator.

We simulate time evolution according to the time-dependent Gross-Pitaevskii equation \cite{Pethick-Smith-2002}, which we write in the form \cite{Caradoc-Davies-2000}\begin{widetext}
\begin{equation}
i \hbar \frac{ \partial \psi(\mathbf{r},t) }{\partial t} =
\left[ -\frac{\hbar^{2}}{2m} \nabla^2 + \frac{1}{2} m \omega^2 r^2 + V_{e}(\mathbf{r},t) + g |\psi(\mathbf{r},t)|^2 \right] \psi(\mathbf{r},t),
\end{equation}
\end{widetext}
where $m$ is the mass of each atom, $\omega$ is the trap frequency, $V_e$ is an external potential, and $g$ is the non-linear coefficient, which is given by
\begin{equation}
g = 4 \pi N \frac{a}{d} \frac{\hbar^2}{m}.
\end{equation}
$d$ is the thickness of the BEC in the third dimension (the $z$ direction), which needs to be very small for the BEC to behave 2-dimensionally.  This is usually ensured by tight $z$ confinement by a harmonic trap that satisfies the condition $\mu < \hbar \omega_z$, where $\mu$ is the chemical potential.
Note that $g$ depends not only on $m$ and $a$, the s-wave scattering length, but also on the number of atoms, $N$, in the BEC.  As the number of particles is contained in the parameter $g$, which is very convenient for the purposes of this paper, the density is normalised according to
\begin{equation}
\int |\psi(\mathbf{r},t)|^{2} d\mathbf{r} = 1.
\end{equation}
In our computer model we represent the BEC wave function on a square area of $16 \times 16$ dimensionless units by a $128 \times 128$ array of double-precision complex numbers.  The dimensionless variables for position and time, $\mathbf{r}^\prime$ and $t^\prime$, are related to the corresponding variables with dimensions, $\mathbf{r}$ and $t$, through the equations
\begin{equation}
\mathbf{r}^\prime = \left( \frac{2m \omega}{\hbar} \right)^{\frac{1}{2}} \mathbf{r}
\end{equation}
and
\begin{equation}
t^\prime = \omega t.
\end{equation}
We solve the Gross-Pitaevskii equation with a fourth-order Runge-Kutta method (a detailed description can be found in ref.\ \cite{Caradoc-Davies-2000}).

\begin{figure}
\includegraphics{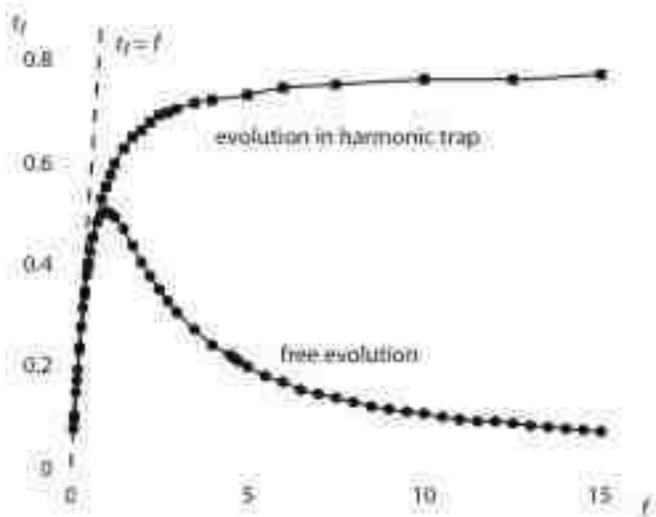}
\caption{Focussing of a BEC by a lens pulse -- a far off-resonant light pulse with a quadratic intensity profile that phase-shifts the BEC locally by $r^{2}/(2 f)$.  The graph shows the numerically calculated relationship between the 'focussing time' parameter, $f$, and the actual focussing time, $t_{f}$, after which a BEC that was initially in the trap ground state is found to be phase-flat.  Analogy with optics suggests that the BEC comes to a focus after a time $f$, i.e.\ $t_f = f$; however, to our surprise we find that this is only the case for small values of $f$.  Our simulations suggest that the relationship shown in the graph holds for values of $g$ in the range $0 \leq g \leq 100$.}
\label{focussing-figure}
\end{figure}

We model the effect of a lens pulse by locally altering the phase of the wave function in the same way in which a lens (or a hologram of a lens) would alter the phase of passing light \cite{Goodman-1968}.  In principle (although it has not been demonstrated for the particular phase profile of a lens), this can be achieved experimentally using an off-resonant laser pulse with an intensity profile proportional to the desired local phase change, whose electric field then -- through the AC Stark effect -- alters the local phase of the BEC.  This 'phase imprinting' method \cite{Dobrek-et-al-1999,Denschlag-et-al-2000} is analogous to optical phase holography.  Interestingly, the time after which the BEC comes to a focus after interaction with such a \textit{lens pulse} is not what one might expect it to be from the optical analogy (see figure \ref{focussing-figure} and appendix).

At this point, it is perhaps worth to briefly discuss the BEC-light analogy in slightly more detail.  The time evolution of an interaction-free, un-trapped, 2-dimensional, BEC according to the Gross-Pitaevskii equation is formally equivalent to spatial propagation of a light beam in the paraxial approximation.  This is, of course, the basis of our specific efforts to reproduce light-like behaviour in BECs.  However, the analogous light beam is of sub-wavelength size, and the paraxial approximation describes the actual propagation of such beams very badly (it does, for example, ignore evanescent waves).  It is therefore not surprising that BECs -- even in the most light-like case of no interactions and no trap -- often behave quite differently from light.

We model local absorption as a straightforward reduction in amplitude; in particular we neglect any interaction between the 'absorbed' and left-over parts of the BEC.  This is an idealisation of a situation which, we believe, can be approximated experimentally in different ways, for example by utilising a two-photon Raman process to couple out (move into an untrapped state) the part of the BEC which lies in the overlap region of two laser pulses  \cite{Edwards-et-al-1999}. To achieve the necessary flexibility, our absorption pattern consists of up to two absorbing cross-hairs (the intensity cross-section is that of a narrow Gaussian, in our simulation often a single pixel wide; for a typical BEC size of 50$\mu$m, this corresponds to lines of light about 500nm wide), as well as an absorbing rectangular boundary with a Gaussian edge profile.

\begin{figure}
\includegraphics{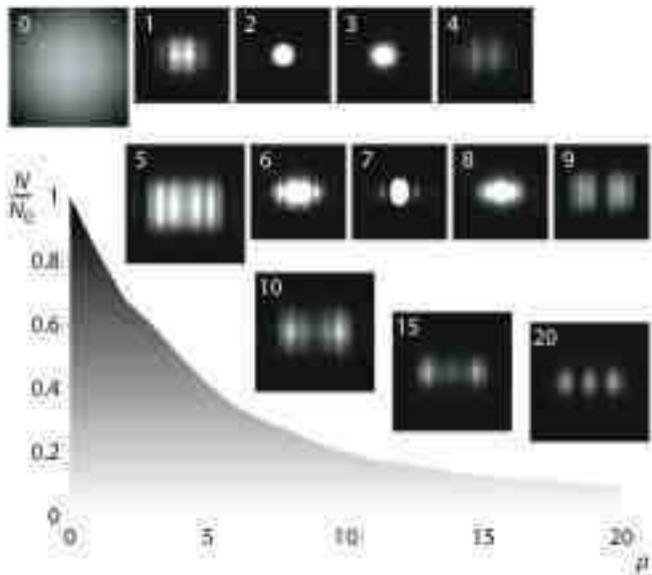}
\caption{BEC after interaction with $p$ lens pulses and absorption events of a light-pulse resonator (the value of $p$ for each frame is shown in the top-left corner), starting with the harmonic trap ground state ($p = 0$), calculated for $g=100$.  The pulses of focal time $f=0.5$ are separated by $L = 0.2$ (in dimensionless time units); the density patterns shown here are calculated a time $L/2$ after the actual pulse, i.e.\ half-time between the previous and the next pulse.  The curve at the bottom shows the fraction of the initial number of particles, $N_{0}$, that remains in the BEC after $p$ periods of the light-pulse resonator.  To demonstrate the loss, the density is not normalised in the frames in this figure; the full scale of grays represents in the frames for $p = 1$ to 4 densities between 0 and 0.2, in the remaining frames between 0 and 0.05.}
\label{first-round-trips-figure}
\end{figure}

Figure \ref{first-round-trips-figure} shows the BEC after the first few light pulses, starting with the trap ground state, through a light-pulse resonator of length $L = 0.2$ and with a focal time $f = 0.5$ (both in dimensionless time units).  We use this as our 'standard' light-pulse resonator.  It will be seen that this behaves in many ways analogous to a stable optical resonator.  Due to the losses, the number of particles changes during every period of the light-pulse resonator, which in turn changes the size of the self-interaction potential in the Gross-Pitaevskii equation, $g |\psi|^2$ (provided $g \neq 0$).  As we keep the integral over the density normalised in our model, the reduction in the number of particles and any corresponding change in the interaction term size is represented as a decrease in $g$.

As long as the value of $g$ -- and with it the potential -- changes, no stationary eigenmode can form.  This means that, for $g \neq 0$, no stationary eigenmode can form before the loss has become zero.  In practice, the number of particles keeps diminishing exponentially, so zero loss can never be reached with any atoms still left.  However, while atoms are still left, the loss can become very small.  Then the BEC becomes very similar to the eigenmode corresponding to the eigenmode for the same value of $g$ in the hypothetical limit of no loss.

In order to calculate eigenmodes for any given value of $g$ in the no-loss limit, we re-normalise the integral over the density after each loss pulse without making a corresponding adjustment to $g$.  This would correspond to either coherently replenishing the BEC after each loss pulse such that the number of atoms remains constant, or to tuning the value of $g$ to 'simulate' this -- neither of these are easily realisable experimentally.  Nevertheless, the procedure gives stationary eigenmodes that correspond to the no-loss limit.

After a number of periods of the light-pulse resonator the eigenmode has formed, that is the BEC is not significantly changing any more over subsequent periods (apart from a uniform factor to the wave function).  As a criterion for reaching this state we demand that the loss due to the aperture varied by less than 0.5\%; this happened typically after about 40 iterations (light-pulse sequences), which takes several minutes to simulate on our desktop computer.  Note that there appears to be an upper limit to the non-linear coefficient, $g_{\rm max}$, above which the BEC does not converge to a stable eigenmode.  The exact value of $g_{\rm max}$ depends on the parameters of the light-pulse resonator.  We determined this value for a number of different resonators and obtained values between $80 \lesssim g_{\rm max} \lesssim 110$.

Once the eigenmode has formed, the absorption due to each absorption event can be very low; we have achieved losses below 0.2\% per absorption event.  However, before the eigenmode has formed the losses can be very high; just how high depends on the starting conditions and the particular choice of shape of the loss pulse.  Note that there is a trade-off between round-trip loss and number of round trips it takes to form eigenmode: thinner cross wires mean lower loss per round trip, but require more round trips.  For well-chosen parameters, of the order of 10\% of the BEC are left by the time the BEC begins to resemble the stationary eigenmode.

\begin{figure}
  \includegraphics{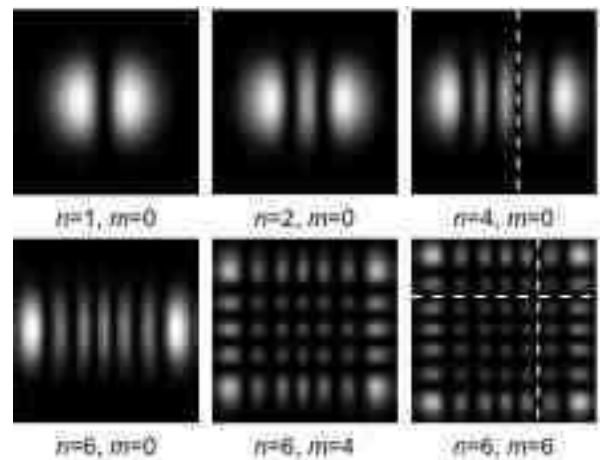}
 \caption{Density patterns of some eigenmodes of the standard light-pulse resonator with a non-linearity of $g = 30$.  By varying the cross-hair position, different patterns -- characterised here by their number of nodal lines in the horizontal and vertical direction, $n$ and $m$, respectively -- emerge, which strongly (but not exactly) resemble optical Hermite-Gaussian modes.  These pictures show the BEC half-way between two lens pulses.  A possible position of the cross-hair (white dashed lines) is shown in the two rightmost frames.  The pictures show only the central $6.25 \times 6.25$ units of the modelled area.}
\label{HG-g30-figure}
\end{figure}

\begin{figure}
\includegraphics{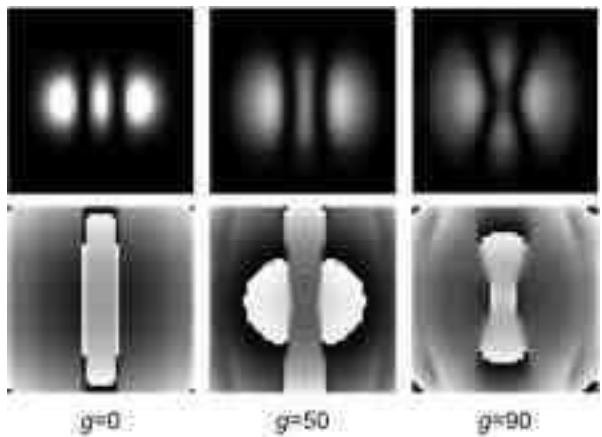}
\caption{Density (top) and corresponding phase patterns (below) for various values of the non-linear coefficient, $g$.  In all cases, the shape of the absorption region is a narrow vertical straight line.}
\label{HG20-figure}
\end{figure}

Figure \ref{HG-g30-figure} shows some of the stable patterns that emerge for a modest non-linearity ($g = 30$, corresponding to approximately $10^3$ atoms of $^{87}Rb$ in a $20\pi$ Hz
trap) by varying the localised absorption pattern -- a 100\% absorptive cross-hair of 1 pixel width placed in crossing node lines (see figure \ref{HG-g30-figure}).  Like all other pictures in this paper (unless otherwise stated), this figure shows only the central $6.25 \times 6.25$ units of the modelled area for clarity, and the grey values represent linear densities, ranging from 0 (black) to 0.2 (white).  The patterns look very similar to Hermite-Gaussian modes, more so for small values of the non-linear parameter $g$, less so for large values (figure \ref{HG20-figure}).  In all cases, the eigenmodes have the tell-tale $\pi$ phase difference between adjacent peaks, just like the optical HG modes.

Repeating these simulations for a switched-off harmonic trap shows that the results are virtually indistinguishable from those obtained with the trap switched on.  This is not surprising as the time scale of effects due to the trap (1 dimensionless unit) is several times larger than the time scale of effects due to the lens pulses (the focal time).  It suggests that a series of lens pulses -- a light-pulse resonator without absorption events -- could act as a novel type of trap.  Indeed, in the absence of absorption events, and to within the accuracy of the Gross-Pitaevskii equation, a series of light pulses acts as a loss-less trap.
It is worth noting that, much like a stable laser resonator 'traps' not only light \textit{waves} but also light \textit{rays}, a light-pulse resonator could perhaps not only trap atoms in a BEC but also hotter or Fermionic atoms.

If the series of light pulses, as well as any other potential (in particular the harmonic trap), is switched off after a stationary eigenmode has formed, the eigenmode evolves freely.  Figure \ref{free-evolution-figure} investigates the size change of the eigenmode during this free evolution; the shape remains almost unchanged.  This property is in analogy to structural stability of the eigenmodes of stable optical resonators, which can be understood in terms of imaging of every plane into every other plane \cite{Forrester-et-al-2002}.

\begin{figure}
\includegraphics{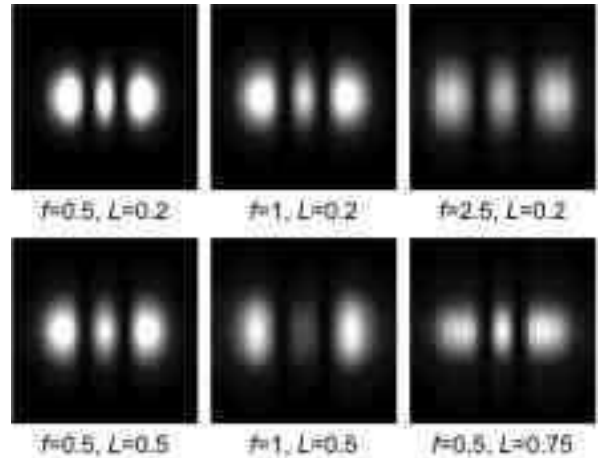}
\caption{Density patterns for the output of different configurations of resonator, calculated for $g = 0$.  $f$ represents the focal time of the lens pulse, $L$ the time between subsequent pulses (both in dimensionless units).  For the particular choice of loss-pulse shape, the loss per period of the light-pulse resonator for these eigenmodes is, from top left to bottom right, 1.8\%, 1.3\%, 5.9\%, 4.4\%, 10.2\%, and 9.7\%.}
\label{HG-different-resonators-figure}
\end{figure}

All the results shown so far were calculated for our standard
resonator with a time between pulses -- corresponding to the resonator length -- of $L = 0.2$, and a lens-pulse focal time of $f = 0.5$.  Figure \ref{HG-different-resonators-figure} shows $n = 1$, $m = 0$ eigenmodes from different light-pulse resonators.  In the optical analogy, altering resonator parameters while keeping the resonator geometrically stable (that is, satisfying $L < 4 f$) results in eigenmodes of a different size; the analogous effect in BEC light-pulse resonators can be seen in figure \ref{HG-different-resonators-figure}.

\section{Conversion into Laguerre-Gaussian-like modes}

To examine the similarities between the eigenmodes of light-pulse resonators and optical HG modes beyond the level of appearance, we investigate here the analogy of a conversion of optical HG modes, which have a rectangular symmetry, into circularly symmetric Laguerre-Gaussian (LG) modes under cylindrical focussing \cite{Beijersbergen-et-al-1993}.  This conversion is due to a fairly subtle effect: the Gouy phase shift, which occurs when HG modes pass through a focus.  It is interesting also because LG modes can contain a higher-charge vortex at their centre.

 \begin{figure}
 \includegraphics{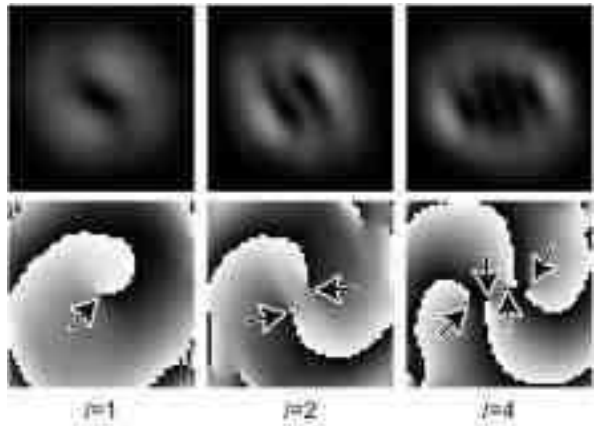}
 \caption{Modelled density (top) and phase (bottom; grey-level representation) after a pair of cylindrical-lens pulses has passed through eigenmodes with, from left to right, $n = 1, 2, 4$, respectively, and $m = 0$, in a BEC with $g = 50$.  The patterns were obtained with cylindrical-lens pulse pairs with, from left to right, focal time $f = 0.23$ and separation $s = 0.35$, $f = 0.23$ and $s = 0.3$, and $f = 0.1$ and $s = 0.11$, respectively.  The phase patterns show the presence of respectively $l = 1, 2, 4$ vortices, situated at positions where all possible phases meet (marked by arrows).}
\label{LG-figure}
\end{figure}

In the optical case, the cylindrical focussing that converts HG modes into LG modes is usually performed with a pair of cylindrical lenses.  In analogy, we use here a pair of cylindrical-lens pulses, i.e.\ far off-resonant laser pulses with an intensity profile that varies quadratically in one direction and which is constant in the perpendicular direction.  Figure \ref{LG-figure} shows some results that were calculated for $g = 50$.  The density is certainly not shaped like a perfect ring; instead, most of the density is contained in a distorted ring.  Also instead of containing a charge-$l$ vortex at the centre, the wave function contains $l$ vortices of charge~1.  The conversion is therefore not perfect, but there are some indications that it is happening.  In other simulations (not shown) we confirmed that this is also the case for different choices of $g$ (including $g = 0$), and that, in the appropriate cases, the shape of the BEC also shows signs of the multiple concentric rings characteristic of LG modes with a radial mode index $p > 0$.

In the optical case, the focal lengths of the cylindrical lenses and their positions need to be carefully matched to the parameters of the incoming HG mode to achieve good conversion into a LG mode.  We tried to similarly optimise the parameters of the cylindrical-lens pulses.  However, the similarity to a pure LG mode of the resulting density was always limited, which is not surprising as the BEC was initially not in a pure HG mode and as the evolution of the BEC is different from that of light.  Finally, optimisation of the pulse parameters was not made easier by the complications in the focussing characteristics of a BEC mentioned earlier (figure \ref{focussing-figure}).

\section{Conclusions}

The emergence of Hermite-Gaussian-like modes in a BEC is a very graphical further illustration of the similarity between BECs and laser light.  However, any experimental verification of our scheme would have to take into account the high losses associated with the absorption events.

Alternatively, the absorption events could be replaced by far off-resonant pulses that locally disturb the BEC in places where the local density of the desired mode is zero, thereby effectively transferring a small part of the BEC in an undesired mode into a variety of other modes, amongst them the desired mode.  No simulations of this scheme have been performed to date, so the magnitude of any loss saving is unknown.

Our results indicate that a series of lens pulses acts as a novel trap.  Although analogy with optical resonators, in agreement with our approximate description with the Gross-Pitaevskii equation, would suggest that such a trap is loss-less, a more detailed microscopic description of the BEC predicts finite losses during the interaction with the lens pulses and when the density gets high due to mechanisms such as 3-body recombination~\cite{Kagan-et-al-1998,Fedichev-et-al-1996}.

As described in this paper, a light-pulse resonator confines a BEC in two dimensions.  It should be possible to achieve confinement in all three dimensions by a combination of two orthogonal light-pulse resonators or similar arrangements.  Alternatively, the rapid fall-off of the intensity away from the focal plane due to extremely tight focussing of the lens pulses, which leads to an approximately quadratic Stark potential along the propagation axis, could perhaps be used to extend the confinement to all three dimensions.

We are currently investigating the BEC analogy to holographic light shaping.  In this way, a BEC can be shaped in two and three dimensions.  This could, for example, be useful in quantum computation with arrays of trapped atoms \cite{Monroe-2002}.  The analogy can even be extended to the extremely versatile volume holograms, whereby the BEC analog of a controllable volume phase hologram -- a far off-resonant light pulse with a controllable time-varying intensity -- might even be easier to realise than the original.

\begin{acknowledgments}
\noindent
JC is a Royal Society University Research Fellow, P\"O acknowledges support from the Royal Society of Edinburgh.  Financial support is also gratefully acknowledged from the UK Engineering and Physical Sciences Research Council (EPSRC).
\end{acknowledgments}

\appendix

\section{Analytical focussing results}

Let us briefly discuss the focussing dynamics in two dimensions. For low densities or small scattering lengths we can assume a non-interacting gas, i.e.\ $g=0$. Using the propagator for the now linear Schr\"odinger equation with a harmonic external potential we obtain the full dynamics of the condensate,
\begin{eqnarray}
\Psi(r,t) &=& \exp \left( -i\frac{r^2}{4} \cot(\omega t) \right) \frac{1+i\cot(\omega t) }{1+i(\cot(\omega t)-1/f)} \nonumber \\
&& \times \exp \left( \frac{r^2}{4\sin^2(\omega t) \left( 1+i(\cot(\omega t)-1/f) \right) } \right),
\nonumber \\
\end{eqnarray}
where we have used the initial state 
\begin{equation}
\Psi(r,t=0)=\exp \left[ {-\frac{1}{4} r^2-i\frac{1}{4 f} r^2} \right],
\label{initial-state-equation}
\end{equation}
This initial state is a ground state whose phase has been altered by a lens pulse of focussing time $f$.  All lengths are in units of $(\hbar / 2m \omega)^{1/2}$ and the energy is in units of $\hbar \omega$. 
From the minimum of the width of the Gaussian $|\Psi(r,t)|^2$ we obtain the actual focussing time 
\begin{equation}
t_f = \frac{1}{\omega}\frac{1}{2} \arctan (2 f).\label{id}
\end{equation}
If $f \gg 1$ we get $t_f = \pi/4$.  In the opposite limit, $f\ll 1$, $t_f \approx  f$, in accordance with figure~\ref{focussing-figure}. 

We can also calculate an analytical expression for the focussing time when the interactions between the atoms are not negligible and the inequality $\mu/\hbar\omega\gg1$ (where $\mu$ is the chemical potential) is satisfied -- the Thomas-Fermi regime. During the focussing the condensate is structurally stable, i.e.\ the shape of the cloud is preserved and only the time and length scales change during the evolution.  We rewrite the Gross-Pitaevskii equation in the hydrodynamic form 
\begin{eqnarray}
&&\frac{\partial\rho}{\partial t}+\nabla\cdot(\rho {\bf v})=0 \label{cont}\\
&& m \frac{\partial {\bf v}}{\partial t}=-\nabla \left( \frac{1}{2}m v^2+ \frac{1}{2} m \omega^2 r^2+g \rho-\frac{\hbar^2}{2m}\frac{\nabla^2 \sqrt{\rho}}{\sqrt{\rho}} \right), \nonumber \\
\label{vel}
\end{eqnarray}
where $\rho(r,t)$ is the density with the velocity ${\bf v}$ defined as 
${\bf v}= (\hbar / m) \nabla \Theta({\bf r},t)$ and $\nabla \Theta({\bf r},t)$ is the gradient of the condensate phase. 
In the Thomas-Fermi approximation we neglect the last term on the right-hand side of Eq.\ (\ref{vel}). The density will now take the form
\begin{equation}
\rho(r,t)=\frac{1}{g b(t)^2} [\mu-\frac{1}{2}m\omega^2 (\frac{r}{b(t)})^2 ],
\label{den}
\end{equation}
where  $\mu$ is the chemical potential at $t=0$.  If we choose the phase to be quadratic in $r$, 
\begin{equation}
\Theta(r,t)= \frac{m}{2\hbar} r^2 q(t), 
\end{equation}
and insert this into Eqs.\ (\ref{cont}) and (\ref{vel}) together with the parabolic density $\rho(r,t)$, we see that the dynamics is completely described by the parameter $b(t)$ if and only if $b(t)$ fulfills 
\begin{equation}
\ddot b=\omega^2(\frac{1}{b^3}-b)
\label{nonlin}
\end{equation}
with $q(t)={\dot b}/b$ and the initial conditions $b(0)=1$ and $\dot b(0)= -1/f $. Equation (\ref{nonlin}) can be solved exactly to give
\begin{equation}
b(t)=\sqrt{1+\frac{1}{2f^2}-\frac{1}{2}\sqrt{ \frac{1}{f^4}+\frac{4}{f^2} } \sin \! \left( 2t+\arctan(\frac{1}{2f}) \right)} 
\end{equation}
The focussing time, defined by $\dot b=0$, is consequently
\begin{equation}
t_f=\frac{1}{\omega}\frac{1}{2} \arctan(2 f)
\label{tf}
\end{equation}

As can be seen from Eqs.\ (\ref{id}) and (\ref{tf}) the focussing time is the same both for the noninteracting case and in the Thomas-Fermi limit. This is a coincidence and holds only in the self-similar situation and in 2D since the radial excitation frequencies are given by $\Omega=2 n \omega$ for the ideal gas and $\Omega=\omega\sqrt{2n^2+2nm+2n+m}$ \cite{Stringari-1998} for the Tomas-Fermi cloud. For $n=1$ corresponding to the lowest breathing mode, i.e.\ focusing, the frequencies are the same.

%\bibliography{ARTICLE}

\end{document}